\documentstyle[12pt,a4]{article}
\begin{document}

\title{ Motivations and Physical Aims of Algebraic QFT}

\date{July 1996}

\author{\bf Bert Schroer\\
Freie Universit\"at Berlin\\
Institut f\"ur Theoretische Physik\\
Arnimallee 14
14195 Berlin\\
e-mail: schroer@physik.fu-berlin.de}
\maketitle

\begin{abstract}

We present illustrations which show the usefulness of algebraic QFT.
 In particular in low-dimensional QFT, when
 Lagrangian quantization is not available or is useless 
(e.g. in chiral conformal theories), the algebraic method is beginning to 
reveal its strength.
\end{abstract}

\begin{center}
 {\bf This work is dedicated to the memory of
 Claude Itzykson}
\end{center}
\newpage

\section{History and Motivation}

One characteristic feature which distinguishes algebraic QFT from the various
quantization approaches to relativistic quantum physics 
 based on classical Lagrangians (canonical, path-integral), is its 
 emphasis on
locality and its insistence in separating local from global 
properties. The
former reside in the algebraic structure of local observables, whereas the
latter usually enter through states and the suitably 
constructed representation
spaces of local observables. 

The idea that the ``Global" constitutes itself
from the ``Local" is of course the heart-piece of classical electromagnetisms
as formulated by Faraday, Maxwell and Einstein. 

Algebraic QFT is more faithful
to physical principles than to particular formalisms. As such it has more in
common with the Kramers-Kronig dispersion relations of the 50's as a test
of Einstein causality, than with the
 post Feynman developments of functional formalisms. Its conceptual
and mathematical arsenal is however significantly richer than the QFT 
underlying
the derivation of the Kramers- Kronig dispersion relations.

When physicists first analyzed the new concepts of standard quantum theory, 
they payed little attention to local structure since their main aim was to
understand the new mechanics. For example in von Neumann's 
early general account
of the mathematical framework of quantum physics in terms of observables and
their measurement, locality and causality did not enter at all. 
Even in the subsequent refinements of Wick, Wightman and Wigner \cite{1}
concerning limitations on the superposition principle, locality was not used.

The idea that, by combining the superselection principle with relativistic 
locality, one may obtain a powerful framework in which global superselected
charge sectors including their (asymptotic) particle structure together with
the statistics could emerge from the more fundamental ``nets" of local
observables, can be traced back to the work of Rudolf Haag in the late 50's
\cite{2}. 

More than a decade before, E.P. Wigner \cite{3} had successfully
demonstrated that it is possible to investigate fundamental issues of
relativistic quantum physics without any recourse to quantization. Here, we
are of course referring to his famous representation theory for the Poincar\'e
-group which ended the long-lasting (but somehow academic) discussion about relativistic
linear field equations.

By the end of the 50's, there also existed an ``axiomatic" framework \cite{4}
(the ``Wightman framework") for point-like covariant fields, which, if enriched
by the LSZ asymptotics \cite{5}, gave a unified viewpoint linking the 1929
Heisenberg-Pauli canonical commutation approach \footnote{We will, whenever 
possible, refer to textbooks or reviews where the content
of historical papers is presented from a viewpoint similar to ours.}
\cite{6} with the 1939 Wigner representation theory under one roof.
\footnote{Behind these elegant schemes which were based on a few principles was
of course the sweat and tears of the pioneers of renormalized perturbation
theory.}

The beginning of what is nowadays called algebraic QFT is usually identified
with the 1964 paper of Haag and Kastler \cite{7}, although there were prior
important developments notably by Borchers and Araki \cite{7}.
Borchers showed, among other things, that the scattering matrix is an invariant
shared by a whole local class of fields (the ``Borchers equivalenc class"),
and Araki demonstrated that the local observable algebras of algebraic QFT
have no minimal projectors (i.e. they allow
no maximal measurements), even for the case of free fields.
In contemporary  mathematical terms they are type
$III_1$ hyperfinite von Neumann factors, a fact which (as will be 
commented on
later) has deep physical relations to such diverse looking issues as the
existence of antiparticles and the Hawking-Unruh effect.
The derivation from the physical principles of local observables
of the properties of charge-carrying fields, including their
statistics and internal  
symmetries,
was an important result of the early 70's and is nowadays referred
to as the DHR theory \cite{7}. Its specific adaption to low-dimensional QFT with 
$d\le2+1$, which leads to braid group statistics, is of a more recent vintage
\cite{8} temperatur physics.

For reasons of completeness we should add the remark that nontrivial superselection
rules (generalized "$\theta$-angles") can also be obtained without local
observables in (global) quantum mechanics. However, in that case,
strictly speaking, one has to
go beyond the Heisenberg-Weyl-Schr\"odinger theory and consider
$C^*$-algebras with finite degrees of freedom which, like the particle on a
circle (the ``rotational $C^*$-algebra") or the rigid top ($C^*$-SO(3)
group algebra), are not simple and hence admit inequivalent irreducible representations.
In all those global cases, the superselection rules are {\it not fundamental} but 
rather result from a {\it physical over-idealization}. Take the example of the
string-like Aharonov-Bohm solenoid. This is of course the idealization of a
finitely extended physical solenoid in Schr\"odinger-theory. It so happens,
that in the singular limit, when the mathematics simplifies, the $C^*$-algebra
becomes the rotational algebra, which is conceptually different from the Heisenberg-Weyl 
algebra.  The fact that there is a single $C^*$ algebra with inequivalent
irreducible representations is 
often overlooked in the differential geometric (fibre bundle) framework which tends
to interpret different representations as different theories, 
and not as superselected sectors of just {\it one} $C^*$-algebra.

Superselection rules in quantum mechanics have more similarity to the ``seized"
vacuum structure (generalized $\theta$-angles) than to the charge superselections
structure of QFT.
Only in local theories with infinite degrees of freedom i.e. quantum field
theories (also  in the thermodynamic limit of lattice theories)
one may encounter fundamental superselection rules as a natural 
consequences of
the local observable structure (and the various globalizations it naturally
leads to). In the present context it is more important to stress the differences 
\footnote{To view local quantum physics as just a particular extension of
quantum mechanics or a quantized extension of classical field theory is in my view
somewhat misleading. It is rather a new realm of physics with novel concepts.}
of local quantum
physics (with entirely new concepts) with respect to quantum mechanics than
some similarities between them.

The assumptions on which the algebraic method is founded can be separated
into two groups.\\
\\
(1) {\sl The local algebraic net properties for the observables.}
 
They are formalized
in terms of a coherent map ${\cal A}$ of finite space-time regions $O$ into von
Neumann-algebras:
$${\cal A} : O\to{\cal A}(O)$$
It is sufficient to define this  ``net" (i.e. family of von Neumann algebras
indexed by space-time regions) for the smallest family of regions 
stable under Poincar\'e-transformations which is the family ${\cal K}$ of ``diamonds"
or double cones. It is generated from double cones centered symmetrically
around the origin by applying Poincar\'e transformations. The local algebras
${\cal A}(O)$ are operator algebras in an auxiliary Hilbertspace. Among the
coherence properties they satisfy is the space-like commutativity (Einstein
causality).\footnote{This {\it locality principle} was, and continues to be
(see also section 5 on chiral conformal field theories) the {\it heart-piece of
the algebraic approach} and local quantum physics.}
 Often the auxiliary Hilbertspace is chosen to be physical the representation
space of the vacuum representation. 

The net of von Neumann algebras lends
itself to two globalizations: the $C^*$-algebra of quasilocal observables
(the inductive limit for families ${\cal K}$ directed towards infinity) ${\cal A}_
{quasi}$ \cite{7}, or the larger universal observable algebra
${\cal A}_{uni}$ which corresponds to an algebraic analogue of geometric space-time
compactification.
space-time \cite{9}.\\
\\
(2) {\sl The global quantum aspects of states} 

Since  the (globalized) net allows for many  inequivalent representations,
one is forced to work with a more general concept of states than that of
Schr\"odinger quantum mechanics \cite{7}. In QFT states are linear positive
functionals on the observable net which satisfy certain continuity properties
with respect to this net. Among the physical state there is the distinguished
vacuum state. Behind the vacuum requirement (or the use of the closely related
KMS temperature states) is the stability idea \cite{7,13}.

At this point it is helpful to alert the reader against potential misunderstanding
of the algebraic approach as some form of ``axiomatics" or ``doctrine" on local
quantum physics. As already in case of von Neumann's concepts of observables, the idea
of a net of local observables is nothing more than a useful ``ordering principle of the mind" in
order to separate a part of reality about which we can safely apply our intuitive
locality and causality principles, from useful additional 
constructions like e.g. charge-carrying
fields with weaker and less immediate localization properties. For the latter  our 
intuition is less
reliable, since it is based more on 
experience with formalisms and less on physical principles.
Whether one extends observable nets or makes them smaller, does not 
significantly influence 
the properties of the complete theory (in as much as in the analogous problem
of Heisenberg's cut  between the system and
the observer in the process of measurement 
the exact position of the cut is flexible as long as there 
is a cut somewhere).

Looking at the dichotomy between observables and states the question
``where are the covariant pointlike fields?" will invariably arise. This
question is best answered through an analogy with differential geometry.
The fields serve as a ``coordinatization" of the local algebras. They play
the role of a kind of pointlike generators, and very different looking systems
of fields may generate the same local algebras. The intrinsic physics resides
in the local net and its physical states, but for explicit model calculations
it is convenient to use fields. On the other hand the structural properties of
the theory are more appropriately analyzed within the net formalism. 
Certain ``coordinates" are however distinguished, e.g. local Noether currents
and charge carrying fields with the smallest dimension (the relevant variables
in the language of critical phenomena) in their respective superselection class.

The
deemphasis on fields in favour of local nets is typical for the algebraic method
as compared to quantization approaches. Once a property has been understood
within the algebraic setting, it is clear that one is dealing with an intrinsic
aspect of the physical system i.e. an aspect which does not depend on the 
formalism by which the theory has been constructed. Not all properties obtained
by applying the various quantization approaches are intrinsic in this 
sense,
as examples will show later on. The notion ``intrinsic" which I use throughout
this article has an analogous significance as ``operational" or ``observable"
in the early days of quantum mechanics when ``Gedankenexperiments" were used for
clarification. Here its meaning is a bit more technical. Properties which can
be reconstructed by just using the physical correlation functions are called
intrinsic and structures which depend on the path of construction and cannot be
unambiguously deduced from the observables (example: local gauge invariance, axial
anomalies) are extrinsic.

Among the important discoveries, which originated from a quasi-classical or
perturbative view about QFT that were successfully incorporated and extended
by the algebraic method, are the Goldstone-Nambu mechanism of spontaneous
symmetry breaking and Schwinger-Higgs  ``screening" mechanism in abelian gauge theories.
In algebraic QFT these ideas are elevated to the level of structural theorem
on equal footing with the TCP and Spin and Statistics Theorem \cite{10,11,12}.

Unlike the Wightman framework, the algebraic method, as previously mentioned,
 does not make any a priori
assumptions on the form of commutation relations between charge carrying fields.
The representation theory (superselection theory) of observable nets is powerful
enough to determine those commutation relations 
(modulo Klein transformations, which are then fixed by invoking locality)
\cite{26}.

Even perturbation theory, if looked upon in the spirit of algebraic QFT, presents
new and interesting problems. Here the recent discovery of the ``microlocal
spectrum condition" \cite{13}, following an idea of Radzikowski,
has been instrumental in the re-surging interest in perturbation theory with
algebraic methods. It is interesting to note that this local form of spectrum
condition was discovered in the quantum field theory in curved space-time,  where
a good substitute for the (in that case meaningless) global vacuum reference
state (in form of a new principle behind the recipe of the ``Hadamard 
condition''
\cite{13} or the ``adiabatic vacuum" prescription) had been particularly pressing.
More comments on these very recent results can be found in section 5.

As far as local gauge theory is concerned, the difficulties which algebraic
QFT had in attributing an intrinsic meaning to an elusive ``local gauge 
principle" have deepened. Comments on this problem can be found in many
papers on algebraic QFT and there are  examples of low dimensional gauge theories
which allow for an alternative formulation 
without local gauge fields and anomalies, thus showing that the local 
gauge 
formulation is at best an option,
but not a principle
\cite{14}. 

A recent criticism of the ``gauge principle" appears in the work of N. Seiberg
\cite{15} in supersymmetric gauge theories. However there remains a
 difference between the Seiberg-Witten interpretation of the
asymptotic freedom in terms the infrared limit 
regime as   phases of the theory 
versus that of algebraic QFT where e.g.
the short distance theory reveals itself as {\sl another} theory which is related to the 
original one in a very interesting way and generically has more
degrees of freedom than the original \cite{16}.
This enhancement of global symmetry ($\simeq$ enhancement of degrees of freedom)
is unfortunately not typical for local gauge theories. It happens abundantly in
two dimensional QFT (example Ising field theory) where the chiral limit has
significantly more sectors than its massive parent theories. The phenomenon is
related to the emergence of nontrivial dual (order-disorder) charges in the
scaling limit whereas in gauge theory it corresponds to short-distance
de-confinement \cite{16}.

With all these differences in interpretation and formalism it should come as
no surprise that algebraic QFT also leads to a different view on the Schwinger-Higgs
screening mechanism. First some historical remarks are in order. In an
interesting early note \cite{17}, Schwinger pointed out the possibility 
that charged matter interacting through electromagnetism,  may suffer a charge
screening accompanied by the appearance of massive photons. 
He did not think in terms of
``condensates" formed by the matter fields. Since he had no rigorous argument
in $d=3+1$, he invented the $d=1+1$ ``Schwinger model" \cite{17}
in order to illustrate
the possibility of a ``screened phase" which he imagined to occur in
$d=3+1$ theories for strong Maxwellian couplings. 

Later Higgs \cite{18} (apparently being
unaware of Schwinger's work) studied scalar electrodynamics and found a
perturbative regime with a condensate of the matter field and a massive photon.
He phrased his findings within the current framework of local gauge theories
and thought about the massive photon as originating from a two-step process;
first a spontaneous symmetry breaking with a Goldstone-boson and then the
photon ``swallowing" this boson and in the process becoming massive. Lowenstein and Swieca
\cite{19} made Schwinger's arguments in the $d=1+1$ model more transparent
as well as rigorous
and showed that there is a chiral symmetry breaking which indeed leads 
chiral to condensates.
Since there is no Goldstone-Nambu mechanism in  $d=1+1$, one notes that this model does not allow 
the aforementioned two-step interpretation. 

As already mentioned before, Schwinger's
ideas were vindicated in  a beautiful structural theorem proved by Swieca \cite{11}.
The theorem assumes that the charge-currents are the sources of a Maxwell
field, but uses no ``local gauge principle" or condensates. For this reason
it is very much in the spirit of algebraic QFT, since the latter are 
less intrinsic concepts. But it seems that the more perturbativly 
manipulative but physically less intrinsic
condensate viewpoint became more popular than Schwinger's screening picture.

Algebraic QFT views the ``would
be" charge carriers not as point-like covariant (and hence gauge dependent)
fields, but rather as semi-infinitly extended locally gauge invariant objects
\cite{20} whose localization regions are space-like cones (in the singular
limit: space-like semi-infinite Mandelstam strings). Although they are locally
gauge invariant, they carry a global gauge charge. 

In the Schwinger-Higgs
phase of scalar electrodynamics these locally gauge invariant extended fields
``condense" into the vacuum (and in this way they improve their localization
properties  by loosing their infrared photon clouds, but only at the expense
of birth of dual magnetic objects with even more severe infrared structures).
 At this
point it would have been nice to illustrate this within the path-integral perturbative
setting and the Faddeev Poppov program including the gauge fixing. But I
was unable to find  an appropriate reference. Apparently the problem of conversion of
the path-integral manipulations into the physically clearer language of operator algebras did
not enjoy much popularity (it 
cannot be done by  differential-geometric techniques).
However the Schwinger model provides at least a nice rigorous operator illustration of these
nonlocal objects in $d=1+1$ \cite{16}.

In sections 2 and 3 I will discuss the physical
aims of the algebraic method in the context of low dimensional QFT where they appear
presently in their clearest form through the braid-group statistics problem.
It has been known for some time, that chiral conformal QFT furnishing the
simplest analytic illustration of braid group statistics. The exchange algebras
carrying this statistics (with the R-matrix structure constants) follow from
the aforementioned principles in a natural way \cite{21}. They represent a
specialization of a more general and abstract \cite{8} concept of algebraic
QFT. 
In recognition of the fact that the richness of chiral superselection rules and
braid-group statistics results from the aforementioned locality and stability
principles (and not just from special ad hoc algebras as Kac-Moody algebras
(current algebras) etc.) we called one of our contributions \cite{21}
``Einstein causality and Artin Braids". However I also should not hide the fact
that my attempts to understand certain global properties as e.g. the ``modular
structure" in the sense of Kac and Wakimoto \cite{52} in the framework of algebraic QFT were less than
successful \cite{53}. Since an understanding of these modularity properties as a 
consequence of the principles (perhaps with the help of the powerful more general
Tomita-Takesaki modular theory) is a physically \footnote{This structure appears
precisely at the place of the ``Nelson-Symanzik symmetry" for the Euclidean
interchange of space and time, i.e. at one of the most profound discoveries of
the 60$^{ies}$.} fascinating challenge (and perhaps besides the braid structure
the only physical reason for being interested in chiral theories), I will certainly try a second time.
It is encouraging that at least a partial result in this direction has been
obtained very recently \cite{54}.

The reason behind this analytic simplicity of chiral conformal QFT is
that on the right or left light cone a genuine interaction is not possible
(a kind of algebraic Huygens principles).
The charged chiral conformal fields should be viewed as the ``freest" fields
which fulfill  physically admissible charge-fusion rules together with the
uniquely affiliated and measurable plektonic statistics (or more precisely R-matrix commutation
relations, since chiral conformal QFT has no Wigner-particle structure).
These new fusion laws, unlike those of compact symmetry group representations,
are not compatible with the Lagrangian formalism. 
In fact, these fields strictly speaking are not even Wightman fields since they
come equipped with source and range projectors. The problem of whether one
can rescue the Lagrangian formalism by combining right and left chiral fields
to local fields is in my view of
an academic nature, since all the computational power resides in the non-Lagrangian
chiral fields. The Lagrangian quantization approach is a useful 
tool whenever it applies, but certainly not a concept required by physical 
principles \cite{22}. 

In $d=2+1$ plektonic theories the charges and their fusion laws become also
attributes of particles. Like in $d=3+1$, a given set of superselection rules
admit many solutions which are distinguished by coupling constants (which
may be thought of as dynamical deformation parameters). There the idea of
freeness may be used to select a distinguished field. The best way to implement
this idea seems to proceed 
by an extension of the Wigner representation theory \cite{23}
as in section 6.
Again it seems that Lagrangians play no useful computational role as soon as
plektonic statistics appears.  The situation for $d=3+1$ is presently less clear.
An refined intrinsic notion of ``interaction" would lead to two options mentioned
later. One uses the relation between the scattering operator $S$ and the modular
aspects of the TCP operator, whereas the other is based on the ``micro-local
spectrum condition". Both attempts are severely limited by the present lack of
understanding localization ($\simeq$ infrared) properties of dual (electric
and magnetic) charged objects. On the other hand there have been many new
insights into the relation of temperature with groundstate QFT (relativistic
KMS condition, fate of superselection rules, fate of spontaneously broken
symmetries). In order not to overburden this article with too many issues, I
will not present these latter developments but rather limit myself to what I
think could be attractive to a pragmatically minded reader with some amount of
curiousity about structural properties of QFT.

We finally would like to attempt to incorporate algebraic QFT into the history
of ideas on relativistic quantum physics. The most influential method of thinking
which has dominated QFT for almost three decades is that of Dirac which is
based on geometric intuition, formal mathematical elegance and the power of
analogies. The ``square root" of the Klein-Gordon operator, the ``filling of
the Dirac sea" (in order to implement the idea of stability in relativistic
quantum physics) and the geometric construction  of magnetic monopoles may serve as
illustrations of the ``Diracian" mode of thinking. 

Even if we nowadays see things
in a slightly different way, nobody can deny that this approach to fundamental
problems has been extremely successful. It appears to me that the discovery
of the electro-weak (phenomenological) theory has been the last real big physics
conquest made in this geometric spirit. All attempts to go beyond this discovery 
(GUT, strings, supersymmetry and probably also most of the more recent 
inventions)
were not met with success. This failure, which by the end of this century 
has developed into a profound crisis of particle physics, QFT, and 
even of adjacent
areas,  may be the result of an exaggerated unbalanced use of the Diracian
approach, rather than an inherent weakness of this method.

On the other hand, there has also been a tradition which emphasizes the importance
of physical principles as the condensated form of all our past experience.
This mode of thinking always confronts actual findings with these principles,
often by the intervention of Gedankenexperiments. It develops its full strength, if this
confrontation leads to an antinomy or paradoxon. One finds this mode of
thinking particularly  in the works of Bohr and Heisenberg, in fact
it was essential for the discovery of quantum mechanics by Heisenberg.
(Schr\"odinger was perhaps a bit more on the Diracian side.)
Also the later
contributions of von Neumann, Jordan, Pauli and Wigner serve as good illustrations.
For the purpose of our presentation, let us call this mode of thinking 
``Bohr-Heisenbergian". A more recent example for this approach is algebraic 
QFT.

I do not know a more appropriate illustration for the difference between the
``Diracian" and ``Bohr-Heisenbergian" way of doing local quantum physics than
the two different ways in which  extended objects  (in particular with
string-like objects) enter local quantum physics. 

The Diracian physicist goes beyond point-like (or
compactly localized) objects for reasons of esthetical pleasure  related
to mathematical consistency and beauty. On the other hand he is able to be
somewhat lighthearted about the underlying physics (viz. the sudden transition
of the string formulated dual model of strong interactions to alleged 
 semi-classical
approximation of quantum gravity of modern string theory).
The development of string theory from the dual model was certainly a
``Diracian" contribution.

A ``Diracian" physicist likes inventions (supersymmetry, superstrings,
quantum groups,...) and sometimes even takes them for discoveries.

Quite differently the ``Bohr-Heisenbergian" physicist generally shuns away from
invention and tends to be less impressed by mathematical beauty  unless it arises
as a result of compelling conceptual insight. If string-like objects are
necessary in order to fully explore a physical principle, he is able to full-heartedly
embrace such objects. 

This was indeed precisely the way how string-like objects entered algebraic 
QFT. At the beginning of the
80's Buchholz and Fredenhagen \cite{4,24} proved a structural theorem 
(following up some problems raised in Swiecas work \cite{11}) which established
the arbitrary thin space-like cone (``semi-infinite string-like") localization
of charge-carrying fields as the a priori best possible localization resulting
from the physical principles of algebraic QFT (together with the mass gap
hypothesis). Later \cite{8,9} it became clear that in low-dimensional theories
this more general localization really occurs in connection with braid group
statistic theories,  e.g. in $d=2+1$ braid group
statistics exhausts this semi-infinite string-like localization in the sense
that any better localization (i.e. compact or point-like) necessarily leads
back to permutation group statistics. In other words only charge-carrying
operators with this semi-infinite extension are able to carry ``plektonic"
charges fulfilling the new fusion laws which are not covered by the
representation theory of compact (internal symmetry) groups. Such fields
do not appear naturally in Lagrangian quantization although attempts have been
made to describe such situations via a Chern-Simons extension of the standard
quantization formalism.

In in the Bohr-Heisenbergian
mode of thinking, the discovery of the connection between localization and
statistics is interpreted as an extension of the Einstein-causality
principle and the stability principle (behind the positive energy condition)
beyond  Lagrangian quantization.

In the construction of these new charge carriers (which result from old principles !) one
has to leave the Lagrangian quantization framework and use the methods offered
by algebraic QFT. One needs many new mathematical concepts and it testifies
to the conceptual strength of the algebraic method, that in developing the
algebraic framework of superselection rules, Doplicher, Haag and Roberts 
\cite{7} whose papers were written a decade before the impressive mathematical
work relating subfactors and ``symmetries" by Vaughan Jones, anticipated
(in a  more limited physical context) some concepts of the latter. This
profound connection was deepened in the context of low-dimensional QFT and
braid group statistics \cite{25,8}.

Looking at the rather long time of  physical stagnation in the
post electro-weak era, one gets the impression that the Diracian method has
been much more fruitful for mathematics than for physics.

Most of the more recent scientific contributions of Claude Itzykson, to whose
memory I dedicate this article, are ``Diracian" with geometrical ideas in the
forefront.

 I remember very spirited conversations on the 
present state and the possible future role of QFT; the last such conversation 
 over dinner in  Berlin in late autumn 1994. With sympathy and respect, it is
easy to enjoy also those conversations which do not lead to an identical
viewpoint.
 Claude listened to my account of the deep crisis in physics 
notably in the area of QFT. Finally he suggested to me not to worry about these 
inevitable developments too much,  and instead enjoy the beautiful
mathematical structure evolving from physical (``Diracian") ideas.
Although Claude  always kept a critical 
distance to the ``physics of everything", he enjoyed certain 
fashionable ideas, because they gave him a chance to work
 on those mathematical structures, which he found so appealing
without having to apologize permanently for doing mathematics 
instead of physics. In that respect times were more lenient than two decades
before, when e.g. one had to present an apologetic behaviour for being
interested in $d=1+1$ ``pathologies".

\section{Standard $d=3+1$ versus Low-Dimensional QFT}

One of the most impressive structural achievements in $d=3+1$ algebraic QFT
is the elaboration of the DHR superselection theory of local nets \cite{7}
which culminates in the DR-theorem \cite{26} on the boson-fermion statistics
alternative and its inexorably linked internal compact symmetry group as
computable attributes of the observable algebra. One very physical way of
understanding its content is in terms of a two-step process:

(i) The derivation of para-boson or -fermion  statistics (mixed Young tableaux)
in the sense of Green \cite{7} from the algebraic net properties and
appropriate state localization properties \cite{7,27}. This first step yields
charge-carrying fields (elements of the ``exchange algebra" or the
``reduced field bundle") with R-matrix structure constants \cite{27}. The latter define a representation of the infinite permutation
group $S_\infty$ and therefore fulfill the relation $R^2={\bf 1}$ in addition
to the Artin $B_\infty$ braid group relations. The assumed state- localization
properties are those suggested by Wigner's analysis of particles as irreducible
positive energy representations of the Poincar\'e group. Not all sectors of
potential interest (charged infraparticles i.e. physical electrons, 
infinite energy quark states) are however  covered by Wigner's theory.

(ii) The mixed Young-tableaux lead to statistical weights which may be
encoded into a multiplicity enlargement of the Hilbertspace and  the
construction of multicomponent fields (number of components = statistical weight).
These tensor fields are bosonic or fermionic and transform under a compact
internal symmetry group which is computable from the structure of the
observable net \cite {26}.

Clearly this kind of insight cannot be obtained from a Lagrangian approach,
rather it constitutes the {\it prerequisite} for the applicability of such an approach.
Indeed, the `para-on' fields, although having the same local observables and the
same S-matrix as the tensorial fields, are more noncommutative (R-matrix instead
of local bosonic or fermionic commutation relations) and less local than the
tensorial fields. They also do not have a (quasi-)classical limit and
hence, unlike their tensorial counterparts, do not fit into a Lagrangian
framework \cite{14}. Therefore the formal requirement to work with fields which can be
thought of as being obtained through a quantization, gives preference to the
tensorial description.

In $d=2+1$ QFT one still can carry out the first step (i) by using the 
string-localization resulting via the BF-theory \cite{4,24} from the mass gap
hypothesis. But whereas in $d=3+1$ this would still lead to exchange fields
with $R^2=1$, this additional relation (which forces the braid-group to 
coalesce to the permutation group) is absent in this case. The statistical
dimensions, apart from the abelian braid group situations, are now non-integer
numbers, and the D.R. construction of encoding these statistical weights into
multiplicities of tensorial fields, i.e. the step (ii), does not work. 

It turns
out that chiral conformal field theories which have much simpler analytic
properties than $d=2+1$ massive theories, lead to the same family of admissible
superselection rules and braid group representations \cite{9}. In both cases
one observes (as a substitute for (ii)) some very fundamental looking differences
from  the standard $d=3+1$ theories. The observable nets now allow for a natural
compactification in terms of a global ``universal algebra" \cite{9}. This 
rather big observable algebra contains ``charge measuring" operators which
turn out to be dual to the ``charge-creating" endomorphisms of the DHR theory.
In addition the new situation is selfdual in the sense that the Verlinde matrix
$S$, which in algebraic QFT just represents the value of the various charges
in the various superselection channels, turns out to be symmetric with
respect to its charge-measuring - charge-creating entries. In the case of
(finite) group symmetry, this only happens for abelian groups; but for the case
at hand, one does not know an internal symmetry concept which ``explains" the
symmetry of $S$ and the non-integer statistical dimensions  in terms of a multiplicity
concepts.

 One observes that the universal observable algebra ${\cal A}_{uni}$
contains also a kind of ``charge-induction" operators  which change charges,
but only in the presence of other charges \cite{14}. These operators, if inserted
between basis elements of the orthogonal intertwiner basis, break the
orthogonality \cite{14} and lead to generalization of the matrix $S$ which
are associated to the representation theory of so-called  mapping class groups.
So it seems that the latter form an important aspect of the ill-understood
new symmetry concept (a ``universal" mapping class group algebra?). The DHR
C$^*$ observable algebras do not contain such interesting central charges.

Presently there is no good comprehension of why in low dimensional QFT the internal
symmetry gets so inexorably linked with the space-time symmetry \cite{14},
whereas in higher dimensions one was unable to ``marry" them in a nontrivial
way  (apart from that physically rather uninspiring ``marriage" via extended
supersymmetry). Witten's topological QFT  in terms of functional
integrals as well as the algebraic approach through localized endomorphism
and their intertwiners (appendix of ref.9) describe e.g. 3-manifolds
invariants in a field theoretic setting but do not really ``explain" 
the aforementioned phenomenon. The $q$-deformation theory of ``quantum groups"
does not cast any light on this problem either; even worse such attempts are inconsistent
with the positivity of quantum theory and their only useful role seems to be 
to furnish admissible 
R-matrices at the roots of unity. The latter also appear in a much more
natural and physical way (compatible with quantum theory) via the 
DHR or the Jones subfactor theory.

Although algebraic QFT denies the existence of a ``local gauge principle" with
a {\it direct} physical meaning, some of its attributed associated structures, as the description
of charge carriers of topological charges in the sense of \cite{7,24}  in
terms of semi-infinite (``Mandelstam-like") spacelike strings or as the appearance
of additional degrees of freedom at short distances \cite{16} (the asymptotic
freedom property of certain gauge theories) are naturally accounted for as an
extended realization of known physical principles. Since
algebraic QFT was introduced with predominant emphasis on {\it structural problems}
of QFT, it is particularly gratifying to note that its methods have proved to
be increasingly useful for the actual {\it construction} of objects as anyons and
plektons which hardly fit into a Lagrangian quantization framework. 

Before
we present some partial results in the next sections, it is worthwhile to take
notice of the fact that the issue of braid group statistics \cite{28} is essentially a
quantum field  theoretical problem for which the recourse to quantum mechanics
(despite particle number conservation)
does not yield a significant simplification.

Consider for example Wilczek's idea \cite{29} of using the Aharanov-Bohm effect
in order to obtain an interaction between pairs of ``dyons". Since this 
interaction is extremely long range, one does not a-priori know the kind of
boundary conditions to be used in the stationary scattering formalism (there
is a corresponding problem in the time-dependent formulation). For physical
reasons the multi-particle S-matrix {\it must} however {\it fulfill}
 the standard {\it cluster property}.
If for $(n+1)$ particles one particle is converted into a  ``spectator" 
(i.e. shifted to spatial infinity), then the $(n+1)$ particle $S$-matrix 
has to coalesce into the $n$ particle $S$-matrix. This consistency condition, although trivially
satisfied for short-range interactions, gives rise to a ``bootstrap" problem
in the present long-range case. The transcription of this problem in terms of
quantum mechanical path-integrals does not seem to solve this consistency
problem with the cluster decomposition.

The $n$-particle boundary condition is only
implicitly defined by this cluster requirement which couples all channels.
Therefore the main advantage of the nonrelativistic limit, namely the decoupling
of these channels through particle conservation is lost, and one may as well face the QFT problem
of plektons right away and then construct the non-relativistic limit afterwards
(since in QFT the cluster properties are safely built in).

The braid group statistics problem has a vague
resemblance with relativistic integrable models of QFT (e.g. the Sine-Gordon
theory) in the sense that, although there is no creation of ``real" particles
(the $S$-matrix is purely elastic), the ``virtual" particle structure is quite
complex.

\section{More on Physical Aims}

After we took notice of the fact that the problem of plektonic statistics cannot
be simply solved by referring first to a ``plektonic quantum mechanics"
(but requires a more careful field theoretic treatment), it is
worthwhile to ask about the physical aims one hopes to achieve with a theory
of plektonic fields.

Most mathematically-minded field-theorist are aware of the rich applications
of standard $d=3+1$ QFT to elementary particle physics and therefore consider
low-dimensional QFT as a kind of ``theoretical laboratory". As a result of
absence of genuine interactions on half the lightcone (Huygen's  principle in
$d=1+1$ leads to chiral factorization and rules out any genuine interaction),
chiral conformal QFT is a kinematical 
family of QFT's which  realizes the complete set of admissible plektonic 
statistics analogous to 4-dimensional free fields.

As a generalization of analytic-structural arguments about 4-point functions
\cite{21} one finds for each admissible family of plektonic statistics (Hecke
algebra plektons \cite{8} and Birman-Wenzl algebra plektons \cite{56}) two 
families for their field theoretic
realizations. In terms of standard terminology in chiral conformal field theory 
they correspond to ``W-models" (observable algebra without continuous group
symmetry) and current-algebra models. There are also good arguments \cite{30}
that the former can be obtained from the latter by the DHR invariance principle
i.e. the application of a conditional expectation through averaging over the
group.
The arguments that these families (apart from isolated exceptional cases)
exhaust all plektonic possibilities are somewhat weaker.

It is physically very profitable to take notice of the fact that these families
(beyond their role as theoretical laboratories) explain and classify the critical
indices of a large class of $d=2$ classical statistical mechanic models in terms
of the statistical phases (related to braids and knots!)  of the associated
noncommutative chiral conformal QFT. Even though some of the Euclidean field
theories, which describe the critical limit of statistical mechanics, may
(modulo renormalization problem) permit an interpretation in terms of Lagrangian
and path-integrals, the noncommutative chiral theory (which carries all the
computational power) is not describable  in such terms (the analytically
continued chiral theory is as noncommutative as the original real time
theory). The aforementioned
analytic determination of 4-point functions based on their plektonic content
does not use quantization ideas. There is the realistic hope that the critical
behaviour of a large class of classical statistical mechanics models (perhaps
even all) may be understood in terms of the associated noncommutative nets and
their braid group statistics which is derived and classified solely on the
basis of the locality principle and the stability requirement mentioned in the introduction.

Although this deep relation of the systematics
of $d=2$ critical indices with the superselection charges of chiral conformal
theories has all the attributes of a miracle, this mystery can be traced back
to the (still somewhat mysterious) discovery of the relation between commutative
(abelian von Neumann algebras) euclidean and noncommutative (causal nets)
real time field theory. The conceptual frameworks including the interpretation of
localization in both theories is totally different, a fact often not 
appreciated by physicist who use the Euclidean formalism for other purposes
(e.g. numerical calculations) than structural investigations of QFT. 
Whereas in $d\ge1+1$ massive theories the analytically continued correlation
functions have maximally two ``physical" boundaries, namely the ``Minkowski-
boundary" (expectation values of noncommutative operators) and the ``Euclidean
boundary" (stochastic expectation values of commuting variables in the sense
of classical statistical mechanics), {\it the restriction of the continued chiral
correlation functions to {\sl any} real one dimensional boundary define a 
positive definite Wightman theory}. This curious property makes chiral theories
more geometric than others and permits the association to Riemann surfaces.
It is (in my view) the origin of the high space-time symmetry (M\"obius-invariance,
diffeomorphism covariance) and the inexorable and mysterious link of space-time
symmetry with internal symmetry in chiral theories (also expected for $d=2+1$).
The Riemann surfaces appear where in standard Wightman theories one had the
BWH analyticity domain of vacuum correlation function. For e.g. genus one, the
original real time QFT lives on one cycle and the ``Euclidean" on the orthogonal
cycle. The position of these cycles is not distinguished and what was called ``Euclidean"
is now as noncommutative as the original theory. In passing, I mention that the
algebraic approach also suggest an interesting intrinsic meaning for two expressions
whose historical origin dates back from a time when chiral conformal QFT was
thought of as separated from the rest of QFT. One is ``holomorphic field" whose
meaning (the literature is very unprecise on such concepts) seems to be ``local
chiral". It would be nicer to use it for the above very curious and distinguished
property of admitting a ``continuous family of physical boundaries" in the sense
of positivity. In this meaning the property of being ``holomorphic" explains the
high diffeomorphism convenience of the chiral theories. The second expression
is ``vertex operators" which at first sight just seems to be another name, for
the
good old ``localizable field". Linking this expression however with the previously
mentioned appearance of ``non-Wightman" field (which carry central source and
range projectors), it becomes a very useful terminology which transcends chiral
conformal QFT.\footnote{I am indebted to D.I. Olive for this suggestion.} 

Plektonic theories in $d=2+1$ are physically more important (physics of quantum
layers), but analytically more difficult than chiral conformal theories. The
main reason is that their charge-carriers have a semi-infinite spacelike string
localization and their superselection structure does not uniquely determine
them, rather they have physical deformation parameters (coupling constants).
These interaction superselected charges may belong to $d=2+1$ Wigner particles, whose
nonrelativistic limits are quasi-particles of condensed matter physics
(which retain their plektonic statistics in the nonrelativistic 
limit.). 

Our
present understanding of plektonic properties already supplies us with two
families of numbers of great potential physical significance \cite{9,14}:\\
\begin{tabular}{lll}
  {\bf Algebraic QFT} &$---$ &{\bf condensed matter physics}\\
 statistical dimensions &$\longrightarrow$ &amplification factors, rel. size
 of degree of freedom\\
statistical phases &$\longrightarrow$ 
&e.m. properties of plektons, fractional Hall effect ?
\end{tabular}

\bigskip

A particle physicist would think of amplification factors as the analogs of
Casimir's multiplicity factors (``$2I+1$") in  cross sections. But such
weights for degrees of freedom also enter (and modify) the BCS relation
between critical temperature $T_c$ and the gap $\Delta$ (at zero temperature):
\begin{equation}
T_c=d^2_\rho ~const.~ \Delta
\end{equation}
 Here $d_\rho$ is the statistical weight of the plekton carrying the $\rho$-
superselection charge. For $d_\rho=1$ the relation is the standard BCS relation.
The quadratic dependence on $d_\rho$ can be made plausible for a hypothetical 
SU(2) fermionic BCS model (linear dependence on the Casisimir eigenvalue 4). Anyons 
(=abelian plektons) have $d_\rho=1$ and change only the statistical 
phases,
but do not amplify the weights.

Note that non-abelian symmetry groups in condensed matter physics can only
appear in an artificial way i.e. by making ad hoc approximations on the 
$U(1)$-invariant many-particle problem. On the other hand, since plektons are as selfconjugate as
the superselection structure of abelian group symmetry, a  transition from
a $U(1)$ liquid fermi phase to a plektonic phase is perfectly conceivable.
It cannot be stressed enough, that the issue of relativistic localization is
only important in the classification of physically admissible plektonic statistics.
As for Fermions and Bosons the results remain valid in the nonrelativistic
domain.
Since, as already mentioned, the relation between the superselection data and the
fields in $d=2+1$ is not unique, we need a principle which selects one realization
among all dynamically possible ones (i.e. a substitute for a Lagrangian). Such
a principle based on the idea of ``freeness" will be abstracted from the
long-range cluster properties of the S-matrix in what follows.

In $d=3+1$ the $S$-matrix fulfills the well-known cluster properties which in
the extreme cluster limit (all pairs become infinitely separated in the sense of
their wave-packet localization) takes the form
\begin{equation}
S \mathop{\longrightarrow}\limits^{cluster}_{limit} 1
\end{equation}
It is believed that the Borchers classes of the various free field exhaust
the possibilities of local interpolating fields with trivial scattering,
however a solid proof only exists for zero mass field \cite{31}. 
If true, this would select
the various standard $SU(N)$ or $SO(N)$-selfconjugate free fields of 
$d=3+1$ QFT as the ``freest" Bosons or Fermions.

For $d<3+1$ this limit generically lead to $S_{lim}\not= 1$. In particular a
more detailed analysis shows \cite{14} that for $d=1+1$ one obtains (as the leading long-distance
contribution) an elastic energy (rapidity) dependent $S$-matrix which fulfills the Yang-Baxter equations
as a consistency condition. 
\footnote{It is a pleasure to acknowlede that A. Zamolodchikov had similar ideas
on an intrinsic notion of ``integrability" (private communication). It testifies
to the naturalness of this idea.}
The only surprise is that such a limiting $S$-matrix
is again the $S$-matrix of a localizable QFT, the starting point of the
so-called ``bootstrap program"
for the the construction of relativistic integrable $d=1+1$ massive models.

In $d=2+1,~ S_{lim}$ seems to be piecewise constant in momentum space \cite{14}
with matrices which jump from one $R$-matrix (representation of a word in the braid-
group) to another as soon as the particle velocities go through a coalescent
degenerate configuration. The cross sections of such piecewise constant
$S$-matrices are expected to vanish and there is some ambiguity in what part
of the asymptotics should be accounted for as a change of inner products for 
in- and out-states, and what part should belong to the proper
$S$-matrix. In any case,
the localized fields which may correspond to such a situation have not yet
been computed. In section 6 we use an extension of the representation method
of Wigner (which is however limited to anyons) in order to shed some light
on the localization- and statistic-properties of ``free" plektonic particles
and fields.

\section{Remarks on Perturbation Theory and Interactions}

Standard perturbation theory aims at the perturbation expansion for the vacuum
expectation values of (time ordered) products of point-like quantum fields.
The most convenient presentation is via Feynman rules, which can be formally derived
either by operator- or by functional methods. The rules for renormalizing those
formal expressions have remained complicated (compared to the simplicity of the
Feynman rules themselves) and an elegant incorporation of renormalization into the operator
or functional-method does not exist (for good reasons, see below). The separation of the local algebraic
aspects from the global state properties can in principle be done at the end 
(assuming that the perturbation series converges in some sense) with the help
of the (suitably adapted) GNS reconstruction. But in practice this remains a very
difficult task and this last reconstructive step is usually left out; sometimes in the erroneous
belief that functional integrals based on classical actions will (at least on
a formal level) always define {\it quantum} field theories.

The algebraic approach to perturbative interactions, even on a formal level,
organizes things in a slightly different way. The first step is the construction
of a net in terms of free fields in Fock-space.
The choice of the Fock-space is determined by what kind of 
free fields one wants to couple. For implementing interaction one uses
 Bogoliubov's operator \cite{32} $S(g,h)$ in Fock-space which has formal unitary
representation terms of time ordered products:
\begin{equation}\label{3}
S(g,\underline h) =T e^{-i\int (W_I(\underline\varphi_0(x))g(x)+\underline\varphi_0(x)
\underline h(x))d^4x} .
\end{equation}
Here $\underline\varphi_o(x)$ stands for the collection of free fields and $W_I$
is a Wick-ordered $L$-invariant (polynomial) coupling. $S$ was used by
Bogoliubov to introduce causal fields in Fock-space:
\begin{equation}\label{4}
\underline\varphi_g(x)=S^{-1} (g,h) {\delta\over \delta \underline h(x)} S(g,h) \bigg|_{h=0}
\end{equation}
Here $g$ is a Schwartz test function which is chosen constant on a large double
cone ${\cal C}$. 

It is then a simple consequence of Bogoliubov's causality property \cite{32}
that $\varphi_g(x)$ is independent of the values of $g$ outside the past
light cone $V_-({\cal C})$ subtended by ${\cal C}$. According to a recent 
observation by Fredenhagen (private communication, to be published)
 a change of $g$ inside $V_-({\cal C}),$
which leaves the constant value in ${\cal C}$ unchanged,
can be implemented by a unitary transformation $U$ in Fock-space:
\begin{equation}\label{5}
 U\underline\varphi_g (x) U^{-1} =\varphi_{g_1} (x) ~~~x\in{\cal C}~~~~~,
g_1\bigg|_{\cal C} =g.
\end{equation}
Since the local net {\sl within} ${\cal C}$ does not change under a common unitary 
transformation (of all of its members),  the fields $\varphi_g(x)$ are 
formal candidates
for ``field coordinates"  of a net in ${\cal C}$.  ${\cal C}$ can
be made arbitrarily  large (the possible divergence of the $U$'s in
(\ref{5}) pose no problem for the net), and hence one obtains a net of algebraic QFT, this 
time not in the vacuum representation but rather in an auxiliary Hilbert space.
States on this net, in particular the vacuum state, lead to physical representations
of the global $C^*$-algebra which are inequivalent to the auxiliary Fock-space
representation. 

Several comments on this procedure are helpful.
In order to get away from the formal perturbative time ordered expressions,
Bogoliubov axiomatized the properties of the test-function dependent unitary
operators $S(g,\underline h)$ in Fock-space \cite{32}. It seems that for the distinction between
different interaction polynomials one needs some smoothness of these $S$ in
terms of their dependence of $\underline f$ and $\underline h$. Even presently it is not known if there
exist such operators in Fock-space which correspond to a given $W_I$, i.e.
whether the Bogoliubov axiomatics has such solutions. 

The second step, namely the construction
of the vacuum (and other) representation is related to the  so-called
``adiabatic limit". This two-step separation avoids the pitfall related to
Haag's theorem \cite{7} ,which can also be circumvented by the conceptually
less attractive
introduction of an infrared cutoff and the construction of the thermodynamic
limit (the adiabatic limit). 
 The chances for existence of the net together with a vacuum state
on it may even improve if one restricts the Bogoliubov method to a smaller 
algebra e.g. the algebra generated by neutral composite fields.
An example is the coupling of a massive vector meson to a conserved spinor current
for which the neutral subalgebra stays renormalizable.

It is interesting to  note, that even on a formal level it is not clear, that
gauge theories fit into this two step perturbative construction. The difficult
point here is that the gauge condition which are necessary for the definition
of operator algebras (e.g. the Gupta-Bleuler condition) are mostly global, even in abelian gauge theories. 
The net formulation however is only consistent with local conditions. 

One may hope that one
should be able to bypass the existence problem of the Bogoliubov axiomatics 
by constructing ``interacting wedge algebras ". 
This hope is based on the recent observation
that the free fields algebras belonging to the (``Rindler" or
``Bisognano-Wichmann" according to upbringing and taste \cite{7})
 wedge region 
can be directly constructed without reference to local fields 
\cite{39}.
This raises the  expectation that there should also be a truly algebraic way to
introduce an interaction for this preferred localization region. Support for
this hope comes from the distinguished representation of the $
S$-matrix in terms of modular conjugations:
\begin{equation}\label{6}
S=J_W \cdot J_{0,W} 
\end{equation}
Here $J_W$ is the modular conjugation of the interacting wedge algebra, whereas
$J_{0,W}$ has the same relation to the ``incoming" interaction-free wedge algebra.
This formula is only a transcription of the representation of $S$ in terms of
the corresponding $TCP=\theta$ operators:
\begin{equation}\label{7}
 S=\theta \cdot \theta_0
\end{equation}
Here we used the Bisognano-Wichmann relation between $\theta$ and $J_W$
which are only different by a $\pi$-rotation around the $x$-axis:
\begin{equation}\label{8}
\theta =U(R(e_x,\pi)) J_W~~,~~ W=tx\mbox{-wedge}
\end{equation}
as well as the action of $\theta$ on $\underline \varphi$ and 
$\underline\varphi_{in}$ 
\begin{equation}\label{9}
\theta\underline\varphi (x) \theta =\cdot\underline\varphi^*(-x)
\end{equation}
hence  $\theta\underline\varphi_{in} (x) \theta^{-1} =
 \underline\varphi^*_{out}(-x)$
\begin{equation}\label{10}
\theta\theta_0\underline\varphi_{in}^* (-x) \theta_0\theta =
 \underline\varphi^*_{out} (-x)
\end{equation}
and hence (\ref{6}) up to a phase factor which may be absorbed into
the $\theta$'s.
Therefore a natural distinguished equivalence transformation between the 
interacting and the
free (incoming) wedge algebra would be given by a ``square root" of 
$S$ (which
is expected to be a kind of field theoretic M\o ller operator). Unfortunately we do not have the vacuum
representation, but rather a representation of the wedge algebra in some auxiliary
Fock-spaces which is related to  the vacuum representation by another
unknown unitary equivalence transformation.
A possible construction of interacting nets in the Hilbertspace of incoming
states based on this distinguished role
of the S-matrix as a modular invariant has been sketched in \cite{40}.

It is very gratifying to notice that the von Neumann algebras of this ``perturbative"
net (i.e. induced by perturbative couplings between free fields) belongs to a 
folium of states (the reference state and all the mixed states obtained by
density matrices constitute a folium) on the local algebras 
(double cones and wedges) which allows
an elegant characterization in terms of a ``microlocal spectrum condition"
\cite{13}. Whereas the usual spectrum condition (vacuum and positive energy
condition) is a global requirement, the microlocal spectrum condition is a
condition which can be formulated directly on the local algebras. It is 
on the one
hand weaker (it characterizes a whole folium of states instead of a unique
vacuum ground state, and it does not distinguish superselection 
sectors), but on the other hand it promises to lead to an intrinsic
local characterization of interactions \cite{13}. 

As mentioned in the introduction,
this microlocal spectrum condition was recently discovered in the quantum field
theory in curved space-time for which the ground state (or vacuum) requirement
becomes meaningless (apart from some special cases which admit a time-like
Killing vector). As far as the curved space-time theories are concerned, the
microlocal property not only led to an understanding of the ``Hadamard recipe"
(and its equivalence to the so-called adiabatic reference state construction)
\cite{13},
but it also permits an incorporation of interacting matter (nonlinear equation
of motions) and a natural extension of the Wick-products and the 
definition of an energy momentum tensor operator.
 The latter is needed in a quasiclassical interpretation of the classical
Einstein equations.

Hence the algebraic method suggests new ways of looking at old problems. In this way
one also hopes to get a better understanding of which difficulties in perturbation
theory already occur
on a local level (i.e. in the construction of nets) and which of those are more
related to global obstructions (i.e. related to states). The free fields
used for the construction of the interaction and the local net become more dissociated
from the physical particle content of the model, so that incorrect pictures about
particles being closely related to fields are avoided.

The quantization
approach via functional integrals is entirely global since
 it is physically meaningless to functionally integrate over the field space
belonging to a local piece of Minkowski-space. In addition such a
  formulation is only meaningful in the framework of a subclass of Euclidean 
theories and one  has to keep in mind
that the Euclidean localization (in the sense of statistical mechanics)  is extremely nonlocal with respect
to the real time localization.

The end of this section is also the natural place to make some more detailed
comments about the most serious problem in contemporary QFT and elementary 
particle physics: the profound crisis of the last twenty five years. Even for
somebody not familiar with the content of the various fashions, the crisis is
already evident from the observation that nothing of any physical relevance
happened after the discovery except of the electro-weak interactions (its
detailed experimental verification). This stagnation is very remarkable in
view of the fact that the first three quarter centuries were so rich in 
discoveries. A glance at the selected papers edited by J. Schwinger 
[``Selected Papers on Quantum Electrodynamics", edited by J. Schwinger, Dover
Publications 1958] or even the later papers edited by Imai and Takahasi
[QFT I and II in ``Series of Selected Papers in Physics] removes any doubt.
Even papers on the most central issues of contemporary physics as the recently
reported (even in the international press) breakthrough in QCD confinement based
on the fascinating  ideas concerning electric-magnetic duality
 seems to be without much computational
consequences and creates the feeling of looking at a faint shadow of a theory.
Since effective Lagrangians are just names for certain momentum space Taylor coefficients
of correlation functions, the relation to local quantum physics of these global
observations remains unclear\footnote{In one of those papers
\cite{15} the author seems to suggest that in order to secure these duality
observations, one may need a different conceptual framework from that in which
these observations were made.}.
More specifically effective Lagrangian seem to be ineffective in exposing important
infrared and localization structures of the charged dual (electric or magnetic)
objects.
This state of affairs which is so starkly different from the aforementioned situation, cries out for
an explanation.

It would be to easy to blame this crisis on the marketing of string theory,
although statements of some of the proponents at conferences 
\footnote{``The string theory allows to understand everything, but compute nothing"
(David Gross's answer to the question of an experimentalist about the
computational aspect of  TOE)} may tempt one to do so. But things are not that
easy. String theory is {\it not} the cause for the crisis but rather one of its most
remarkable manifestations. In my view the roots of the present crisis can be
treated back to the aftermath of the early renormalization theory, i.e. to one of the
most important cross roads of modern physics. The infinities which appeared in
QED are not ``intrinsic", they are rather the result of the repair job one has
to perform on a slightly incorrect ``quantization" (Lagrangian) starting point.
The idea of Kramers concerning the distinction of physical from ``naked" parameters
(the self-energy problem) which came from the classical Poincar\'e - Lorentz
electron theory was extremely important in order to disentangle the physical
(renormalized) answers \cite{44}. However these ideas becomes somewhat of a burden
(like many good ideas) if misread as a kind of parallelism between classical
and quantum field theoretical ideas about the electron. In QFT particles cannot be
introduced in addition to fields, but they  are consequences of local relativistic
fields. Wigner's characterization of particles as irreducible representations of
the Poincar\'e group is already an inexorable part of the unitary representations
coming with quantum fields or algebraic nets. Whereas in the classical case one
had to add the particle structure via those partial models of Poincar\'e and
Lorentz, in local quantum theory there is no such possibility because it is already 
there. Therefore there should be an alternative formulation which allows to
avoid those infinities. Indeed the ``split point method" first used by Schwinger
and Johnson can be used to accomplish this. As one can define free field Wick
products instead of using (global) frequency ordering also by local limits
through formulas involving point split products, one may similarly define composite
normal product fields which enter the field equations as the interacting terms.
An illegitimate interchange of the limit with the additive and multiplicative 
components of such a normal product expression will bring back the infinities. Since the
interaction graphical rules are not as elegant (there are several types of
two-point functions and not just one type as the time ordered propagators in
Feynman's scheme) such renormalization schemes never appeared on a sufficient 
general level in the literature; only in simple two-dimensional models as the
Thirring or Schwinger model they have been used. In general it is much faster
to compute with Feynman rules and repair the  illegitimate start by
regularization and renormalization.

The wrong turn came about by interpreting the subtle and deep relation between
classical statistical mechanics and noncommutative QFT too naively. Whereas
statistical mechanics has natural physical cutoffs, this is not the case in
QFT. In order to appreciate this remark without prejudices, it is helpful to
take a look at the fascinating history of attempts at relativistic {\it nonlocal}
theories and at the notion of elementary length. In the 50$^{s}$ it was
thought that by putting form factors into the interaction Lagrangian one could
avoid the above infinities and encounter a less singular behaviour in correlation
function. But it was soon realized (Kristensen) that through
their iteration such form factors  have a disastrous effect on macro-causality; their presence
wrecks the physical interpretability of the theory. Later Lee and Wick made the
proposal to change Feynman rules by pairs of complex conjugates poles (which would
also ``soften" the light cone singularities) again with the same unacceptable
result of a-causal precursors as was shown later.
The subtle relation between euclidean theory (for which certain cutoffs have
physical meaning) and QFT holds only after removal of the cutoff. This is also
in agreement with the previous remark converning the complete relative nonlocality
between localization concepts in statistical mechanics and QFT.
 If a cutoff in relativistic QFT would be more
than a formal intermediate device without any intrinsic physical significance,
this would have been a truly sensational event. The increasing number of
nontrivial massive local QFT's (Ising QFT, Sine-Gordon QFT,...) which have been
constructed via the nonperturbative formfactor program (starting from factorizing
elastic S-Matrices) have significantly reduced the space for folkloric ideas on
renormalization. More specifically the following ideas turned out to be prejudices
\footnote{The idea that theories which are ``finite" in the sense of Feynman
perturbation theory may also be physically preferred (supersymmetry, string theory)
results from this prejudice.}

(i) quantum gravity has to be invoked in order to make QFT well-defined.

(ii) Light cone singularities which go significantly beyond the canonical behaviour
threatens the existence of quantum fields.\\
Anybody who believes that a cutoff has a physical significance in a real time
QFT is kindly invited to modify one of those low-dimensional models by the implantation
of a relativistic cutoff {\it without wrecking} the physical {\it interpretability}
of the model. He will soon get an impression of the scope of such a problem.
 All structural investigations of QFT tell us that without Einstein
causality the basis for the physical interpretation of the theory is lost e.g.
there would be no basis for the validity of cluster properties,  scattering
theory etc.. Time and again we had to learn that notions like ``a little bit
nonlocal" (as well as a ``little bit nonunitary")  are as nonsensical as
``a little bit pregnant", at least within the accepted framework of QFT.
 Only if one goes completely
outside standard QFT into the direction of {\it noncommutative space-time} one finds
a chance for replacing causality by some new structure \cite{Doplicher}
whose consistency is presently under investigation.  

It is interesting to note that for certain families of lattice theories, cluster
properties and hence the existence of multiparticle (``multi-magnon") scattering
states can be shown; but those are of course not relativistic and moreover the
proofs are {\it much} more involved than the corresponding derivations in local
QFT. Needless to say that my critical remarks naturally also include one of my
own papers: I nowadays consider my contribution on the 4-$\varepsilon$ expansion
to the enhancement of physical knowledge as being smaller than any preassigned
$\varepsilon$ \cite{50}.

I will not follow here the path which leads from the ``original sin" of neglect
of the ``intrinsicness" of concepts in QFT to the unscrupulous treatment of
established physical principles in string theory which in my view the path into the
present crisis.

Rather I would like to mention some modest successes outside this way of 
thinking. Closely related to the ambitious program of finding non-Gaussian
fixed points of the renormalization group is the more modest program of
discovering the physical concepts which determine the spectrum of admissible
critical indices in $d=2$ (i.e. the generalization and extension of Onsagers
work on the Ising model). Starting from Kadanoff's ``Coulomb Gas Representations"
of critical indices for specific family of models and going through the
discoveries of Belavin, Polyakov and Zamolodchikov of the minimal models
(associated to the Virasoro algebra), this program has now reached a certain
conceptional depth through the recognition of critical indices as being
the numerical values of superselection charges related to the ``statistical
phases" of algebraic QFT. The latter relates the spectrum to braids and knots.
It seems that the issue of braid group statistics is susceptible to a complete
classification, at least in ``rational" chiral theories (the analogy with the
classification of finite depth Jones subfactors is more than superficial).

The modern point of view which analyses chiral conformal QFT directly from the
local field theoretical principles (without the intervention of global 
Fourier algebras
as the Virasoro- or Kac-Moody-algebra) remained remarkably close to the QFT
framework in which the present author and collaborators came across conformal
QFT in 1974, an observation which I already elaborated in ref.\cite{14}.
My construction in 1974 \cite{56} of the chiral energy momentum tensor as a
``Lie-field" was inspired by older work on Lie fields starting 1961 with Greenberg
and culminating in the 1967 work of Lowenstein \cite{57}. These Lie fields in
present terminology are (not necessarily {\it conformal}) W-fields, i.e. conformal
causal fields for which (modulo a c-number Schwinger term) the commutators close.
The case of a one-component scalar field was most extensively studied \cite{57},
but the examples, although structurally close to current algebras (bilinear in
free fields), were physically not as interesting as the chiral energy momentum
tensor \cite{56}, not to mention the more recent development on W-algebra.
With the additional knowledge of the differential identities \cite{56} between
fields and current (corresponding to the modern Knisknik-Zamolodchikov identities) and the
old conformal decomposition theory\footnote{For a comparitative study of the old
versus the new approach see ref.\cite{14}.}, the question arises, why was the
richness of chiral conformal field theories (minimal models etc.) not seen 
already at the beginning of the 70's? Partly the reason was in my opinion,
that the confidence with which physicist investigate low dimensional QFT
nowadays  was lacking at that time. Contributions in low-dimensional field theory were often
``buried" in conference reports since they did not enjoy much appreciation.
Another perhaps even more important reason is that a systematic mathematical
study of representation of infinite dimensional algebras was not yet available.
Although mathematicians use more global methods (Fourier coefficients of T's
and j's), the (global) Verma module approach to unitary representations was
important also for ``local quantum physicist" since it generated the missing
mathematical confidence. The rupture between the ``old" physically oriented
and the ``new" geometrically inspired QFT (often more the result of neglect of
the past rather than deep differences in philosophy)
occurred during the second half of the
70's. This in my view unique phenomenon in the history of 20$^{th}$ century physics is one
of contributing factors to the present crisis \cite{14}.

Recently a clever strategy of ``perturbing" chiral conformal QFT has been 
found by Zamolodchikov. Here ``perturbation" is not taken literally (otherwise
one would have to handle terrible infrared divergencies), but rather serves as
the starting point for consistency arguments which select the ``integrable"
massive representatives.
These representatives are probably identical to the theories which correspond to
the large distance limit of the S-matrix in the sense of section 3.
There remains the interesting question of the number of massive superselection
classes (according to section 3 equal to the number of factorizable models)
which correspond to one conformal limit. For example it is known that the
conformal Ising model allows for two different integrable perturbations and
there are indications that these two possibilities are already preempted through
two different interpretations of the Ising partition functions
\footnote{I am indebted to A.A. Belavin and A. Fring for discussions on this point.}.
These are open structural problems where one expects algebraic field theory to contribute.

It is interesting to note that Zamolodchikov's \cite{46} construction picture
runs opposite to the
renormalization group extrinsic approach. Our starts in a completely intrinsic matter
from a classification of rather simple theories (as sketched above) and moves
away towards more complicated theories by applying structural self-consisting
arguments on relevant ``perturbing" fields which are chosen from the list of
composite fields of the chiral theory. It seems that for practical purposes
the way from simple models to more complicated ones is more constructive than the renormalization
group approach. The latter is too structureless for (I have not seen any 
construction of a non-Gaussian fixed point except the ill-fated $4-\varepsilon$
expansion). But on the other hand the algebraic QFT point of view (and 
Zamolodchikov's approach, which is very close) is presently limited to $d=1+1$, even
though it uses concepts which are quite general.  There is no reason to believe
that the superselection structure will determine critical indices in $d>1+1$.

Some of the remarks made in this section should be viewed as a qualification
of my critical comments made at a round table discussion "Physics and Mathematics"
[XI$^{th}$ International Congress of Mathematical Physics, International Press,
Daniel Iagolnitzer Editor].

\section {Duality- and Modular- Properties  of Chiral Field Algebras}

It has been known for some time in algebraic QFT, that certain nontrivial inclusions
of observable algebras in the vacuum representation contain valuable informations
about superselected charges. In the following we will describe three different
type of inclusions.

(i) {\it The ``corona-inclusion" [III. 4.2 of ref. 7] in $d=3+1$ massive QFT with a mass gap} 
\begin{equation}\label{11}
 \pi_0({\cal A} (K_1) \vee{\cal A}(K_\infty)) \subset    
 \pi_0({\cal A}({\cal C}))'.
\end{equation}
Here $K_1$ is a centrally located-double cone which is surrounded by a 
double-cone-ring
${\cal C}$, and $K_\infty$ is the ring-like wedge region consisting of all 
points which are space-like relative to ${\cal C}$ and $K_1$. With other words
$K_\infty \cup K_1$ is causally disjoint from the ring ${\cal C}$.
 The operator notation is standard: $\pi_0$ is the vacuum representation  and the
upper dash denotes the commutant of an operator algebra. 
Note that the inclusion is proper, i.e. there is an obstruction against
equality. Haag duality can only be recovered by extending the algebras
(in the case of the algebra localized on the corona this presumably can be
achieved by defining a suitable halfspace radial disorder operator).
In case of the existence
of only a finite number of DHR superselection sectors, the DR theorem tells
us that this inclusion is equivalent to that defined by the fixed point
algebra under a ``double" of a finite group $G$ (${\cal F}$ is the DR field
algebra):

\renewcommand{\theequation}{\mbox{\arabic{equation}a}}
\begin{eqnarray}\label{12a}
\pi_0({\cal A}(K_1))\vee \pi_0({\cal A}(K_\infty))&=&
({\cal F} (K_1)\vee{\cal F}(K_\infty)) ^{G\times G}  \mid {\cal H}_0\nonumber\\
\pi_0({\cal A}(C))' &=& ({\cal F}(K_1)\vee{\cal F}(K_\infty))^{Diag(G\times G)} 
\mid {\cal H}_0
\end{eqnarray}

\setcounter{equation}{11}
\renewcommand{\theequation}{\mbox{\arabic{equation}b}}
\begin{equation}\label{12b}
({\cal F}(K_1)\vee {\cal F}(K_\infty))^{G\times G} \mid
 {\cal H}_0 \subset ({\cal F}(K_1)\vee {\cal F} (K_\infty)) ^{Diag(G\times G)}
\mid {\cal H}_0
\end{equation}
\renewcommand{\theequation}{\mbox{\arabic{equation}}}

In other words the DR group symmetry of $d=3+1$ charge-carrying fields is 
already visible
on the level of neutral observables through (11). The equality holds if
instead of observables we use field algebras, in particular for free fields.

(ii) {\it The ``quarter circle inclusion" of $d=1$ chiral conformal QFT.}
\begin{equation}\label{13}
\pi_0({\cal A}(I_1) \vee \pi_0 ({\cal A}(I_3)) \subset \pi_0 ({\cal A} 
(I_2)\vee{\cal A}(I_4))'
\end{equation}
Here $I_i,i=1...4$ are four intervals obtained by equipartitioning the circle
(i.e. the localization space of chiral conformal theories). This inclusion is
 selfdual\footnote{It is the chiral analogue of the corona inclusion.}
in the sense that it is equivalent to the inclusion obtained by exchanging $I_1$
and $I_3$ with $I_2$ and $I_4$. Therefore it is not surprising that, apart from
finite abelian groups (we again restrict our interest to the case of a finite
number of superselection rules), the ``quantum symmetry" of superselection sectors
of chiral theories is not covered by groups. 
This is the case which we will treat in this paper.

The theories with a mass gap $d=2+1$ fulfill a similar formula \cite{9} with
the intervals being replaced by spacelike cones.\\
\\
(iii) {\it The ``e.m.-corona. inclusion" for $d=3+1$ Maxwellian-like free fields.}\\
 The formula is the same as (11) i.e. 
\begin{equation}\label{14}
\pi_0({\cal A}({\cal C})) \subset \pi_0 ({\cal A}({\cal C}'))'
\end{equation}
but now ${\cal A}({\cal C})$ is the algebra generated by zero mass neutral spin 1 fields
(electromagnetic fields) i.e. the obstruction is characteristic for 
$m=0$ (and $s\ge 1)$. In particular the chargeless free electromagnetic
fields leads to a proper inclusion \cite{7} without nontrivial superselection
sectors which one could blame for this duality obstruction.
The equality sign can be formally recovered by using indefinite metric
gauge potentials, but
the use of unphysical gauge fields in indefinite metric spaces with BRST conditions
is against the spirit of algebraic QFT.
The latter prefers to deal directly with the cohomological problem hidden behind the
proper inclusion instead of artificially removing it at the expense of
introducing  vector potentials.\footnote{Although vector 
potentials can be avoided in all structural arguments of algebraic 
QFT, they do enter in the Lagrangian quantization approach and in the 
intermediate step of the ensuing standard perturbation theory of physical
observables.} For more details on this inclusion see ref. [40]. It is
believed \cite{7} that the existence of nontrivial electric and magnetic charges (i.e.
Maxwell-like interactions) will remove this obstruction against Haag duality.
It is interesting to note that the order-disorder duality structure in low
dimensional QFT does not permit the simultaneous absence of order and disorder
charges.
Due to the unsolved infrared problems of algebraic QFT, this subject
has not yet been thoroughly investigated in $d\ge 3+1$ theories.

I believe that a local algebraic approach to electro-
magnetic duality (i.e. between $F_{\mu\nu}$ and $\tilde F_{\mu\nu}$ as  in the
Seiberg-Witten models) and in particular the construction of the e. and m. charge-
carrying operators which are responsible for this duality, should start from these
observations. Superselection rules arising from $F$ and $\tilde F$ are
expected to be very peculiar and related to the problematization of the notion
of ``magnetic fields" and quark confinement. The charge carriers of such dual
charges must be essentially different from those in the DHR theory. The 
localization properties are expected to be so weak that standard covariant field
description are presumably not applicable. We will not pursue this matter here.

We now return to the quarter circle situation (ii).
In that case one proves the following theorem.\\
\\
{\bf Theorem 1} \cite{33} {\it The vacuum representation of the abelian current algebra
(multicomponent Weyl algebra) violates Haag duality for the quarter circle
inclusion and fulfills instead the following ``lattice duality":}
\begin{equation}\label{15}
A_L(I_1\cup I_3)'={\cal A}_{L^*} (I_2\cup I_4)
\end{equation}
The von Neumann algebras on both sides are generated from Weyl algebras over
real test functions of subspaces of $C^\infty(S^1)\otimes V$. Let ${\cal S}_L
(I_1\cup I_3)$ denote the subspace consisting of real test functions which
are constant in $I_2$ and $I_4$ and fulfill the condition
$f(z_2)-f(z_4)\in 2\pi L~,~ z_{2,4} \in I_{2,4}$ where $L$= even lattice in
$V$. Then:
$${\cal A}_L(I_1\cup I_3) =Alg\{ W(f),f\in {\cal S}_L(I_1\cup I_3)\}$$
and similarly for ${\cal A}_{L^*} (I_2\cup I_4)$ with $L^*$ being the dual
lattice to $L$.

Note that only in the case of a selfdual lattice one obtains the standard 
form of Haag duality. 

The content of the theorem and its proof becomes clearer if one
maximally extends the multicomponent Weyl algebra. Here ``maximally" means that no
further (bosonic) local extension is possible. It is well-known that such maximal
extensions correspond to even lattices in the sense that the functions $\underline f$
in the Weyl generators $W(\underline f)$ fulfill quasi-periodic lattice boundary
conditions on $S^1$ instead of being periodic. This more general affine set of
functions still maintains the causal net structure, but the extended net
${\cal A}^{ext}_L$ has a finite number, namely $|L^*/L|$ superselection sectors,
instead of (continuously) infinitely many. Collecting the results in
 terms of  theorems we have the following:\\
\\
{\bf Theorem 2} \cite{33} {\it The maximally extended observable net 
${\cal A}_L^{ext}$ fulfills the
following extended duality relation}
\begin{equation}\label{16}
{\cal A}_L^{ext} (I_1\cup I_3)'=\pi_L({\cal F}_{L^*}^{(0)} (I_2\cup I_4)).
\end{equation}
Here the left hand side is the commutant of the $L$-extended observable algebra
in its natural $\pi_L$-vacuum representation. The right hand side is the
field algebra restricted to  the charge-anticharge split (total charge
neutral, indicated by the (0)-superscript) subalgebra and projected onto the
vacuum sector.

The proof of the second theorem is based on the first one. Both proofs are
easy and natural if one adapts the framework of Buchholz, Mack and Todorov
\cite{34} and will not be given here.
This extended duality is the prerequisite  for a geometric form of the Tomita-Takesaki
modular theory  for the quarter circle situation. Here we will only
present an application of that theory to the case of {\it no} nontrivial superselection
rules (i.e. for selfdual lattices or for the CAR fermion algebra).
Let us take the special case of chiral CAR algebras.Then we have:\\
\\
{\bf Theorem 3} {\it (based on Araki's analysis of CAR algebra \cite{35}):\\
Let $U(t)=e^{iht}$ be a one parametric group of diffeomorphism acting as
Bogoliubov automorphisms on the wave function space underlying the CAR
$(S^1)$ $C^*$-algebra. Then there exists an associated $\beta$-KMS state.
If the diffeomorphism has a finite number of fixed points, then this KMS state
has a natural factorial $III_1$ restriction to a subalgebra. }

In this way
one expects to obtain a geometric modular theory for this subalgebra and a suitable
 state, i.e. one hopes
there exists a state which together with the subalgebra has the given diffeomorphism
group as the Tomita-Takesaki automorphism group.

Instead of showing how a proof can be abstracted from Araki's work, we mention two
physically interesting special cases of diffeomorphisms.

$\alpha$) $z\to Dil(t)z$ as the diffeomorphism.\\
In this case the situation of the theorem coalesces for $\beta=1$ with the
modular theory of the semi-circle CAR $(\cap)$ 
algebra  and the vacuum state vector. For $\beta\not=1$ the restricted $KMS$ state
belongs to another folium of the simple $C^*$-algebra CAR $(\cap)$ different
from the vacuum folium.

$\beta$) $z\to T^{-1}_2~Dil(t)T_2$, with $T_2:z\to z^2$ covering of $S^1$\\
This diffeomorphism has four fixed points and belongs to the quarter-circle
situation. Instead of the semi-circle CAR algebra one now considers
 the one belonging to $I_1$ and the
state vector is not the vacuum but it is rather defined implicitly by the GNS construction within
Araki's CAR theory If the vacuum folium contains a $\beta=1$ KMS state for this
diffeomorphism group acting on CAR $(I_1)$, then the state must be the so
constructed one. So in order to maintain a geometric modular theory for 
general diffeomorphisms, one should allow reference states which are different
from the vacuum (and outside the vacuum sector).

To recover Haag duality in its original form \cite{7}, we must construct
the field-algebra ${\cal F}_{L^*}(S^1)$ for the extended  current algebra model. 
This algebra contains all localizable charge carries and it fulfills
the following twisted Haag duality.\\
\\
{\bf Theorem 4} \cite {36} ~~~~${\cal F}_{L^*} (I) ={\cal F}_{L^*} (I')^{tw}$\\
\smallskip
Here the superscript on the right hand side denotes the ``twisted" commutant.
In our case of $Z_{|L^*/L|}$-abelian models, the twist is a generalization of the
fermionic twist namely  a suitable 
``square root" of the rotation $e^{-2\pi iL_0} $\cite{35}.

Up to now we have been dealing with abelian (``anyonic") representations
of braid group statistics (level 1 multicomponent current algebras). How does
one construct plektons? Since a ``quantum symmetry" generalizing the
compact group symmetry
does not (yet) exist, a conservative substitute for the previously discussed
abelianization of ``para-ones" which leads to bosons and fermions 
(i.e. the DR theorem) would be to try to
``anyonize" plektons. In the following we will sketch such an
attempt. 

Let us start again from the multicomponent current algebra. It is
well-known that the current algebra in the Weyl form defines an orthomodular
functor \cite{37} from real Hilbert subspaces ${\cal H}_R$ of the multicomponent
complex wave function space ${\cal H}=L^2 (S^1)\times V$ into von Neumann
subalgebras ${\cal N}$ of $B({\cal H}_{Fock})$.
$$
{\cal H}_R\longrightarrow {\cal N} ({\cal H}_R)\eqno{(18a)}$$
$${\cal H}_R' \longrightarrow {\cal N}({\cal H}'_R) 
={\cal N}({\cal H}_R)'\eqno{(18b)}$$
Here ${\cal H}_R'$ denotes the {\it symplectic complement} 
of ${\cal H}_R$ with the
symplectic form  $\sigma(\underline f,\underline g)$ 
being given by the imaginary
part of the scalar product.
\setcounter{equation}{18}
\begin{equation}\label{19}
\sigma (\underline f,\underline g):=Im (\underline f,\underline g)
= {1\over 2\pi}~\int^{2\pi}_0 < \underline f (\theta),\underline g' (0) > d\theta
\end{equation}
${\cal H}_R$ is called standard if ${\cal H}_R\cap i{\cal H}_R=0$ and
${\cal H}_R+i{\cal H}_R$ is dense in ${\cal H}$. In that case ${\cal N}
({\cal H}_R)$ turns out to be a {\it factor in standard position} with the vacuum
state vector in Fockspace being a cyclic and separating reference vector.

If the real subspace consists of real multicomponent functions localized in
an interval $I\in S^1$, the T.T.- modular theory is geometric and the modular
automorphism is the dilation which leaves the endpoints fixed, whereas the
modular conjugation is an antilinear ``reflection" of the algebra ${\cal N}
({\cal H}_R(I))$ onto ${\cal N}({\cal H}_R(I'))$. Both modular operations are
the functorial image of operators in ${\cal H}$. Note that {\it localized} real
subspaces are automatically standard. This is the wave function pre-version of the
Reeh-Schlieder theorem for the localized Weyl-subalgebras acting on the
vacuum. It also may be viewed as a property of coherent states in 
Fock space
affiliated with localized real wave functions. 

The previous maximal local extension by an even lattice amounts to an
extension of this functor to
real ``affine"  spaces":
\begin{equation}\label{20}
{\cal H}_R^{aff} \subset {\cal H} \times L
\end{equation}
Here $L$ is an even lattice which appears in the quasiperiodic boundary conditions
of the real multicomponent
wave function. The extended symplectic bilinear form is
\begin{eqnarray}\label{21}
&\sigma (\underline f,\underline g)= {1\over 4\pi} ~\int^{2\pi}_0(\langle
\underline f(\theta),\underline g' (\theta)\rangle - \langle f'(0),g(\theta)
\rangle )d\theta +{1\over 4\pi} (\langle f(0) \cdot g(2\pi)\rangle - \langle
f(2\pi), g(0)\rangle) &\nonumber\\
&f(2\pi)-f(0)=2\pi\ell_f,~~g(2\pi)-g(0)=2\pi\ell_g&
\end{eqnarray}
where $\ell\in L$ is a lattice vector which is determined by the quasiperiodic
boundary conditions.

So the original Hilbertspace ${\cal H}$ becomes augmented by a lattice and the
extended functor defines a map from real affine subspaces to von Neumann
algebras which are subalgebras of the charged field algebra. The latter is
a kind of crossed product between the previous neutral observable algebra and an 
abelian group algebra representing $L$. The localized subalgebras ${\cal A}_L
(I)$ correspond to rather complicated real affine wave function subspaces
\cite{34} which are not linear subspaces.

 In this sense
the aforementioned maximally extended multicomponent current algebras are still
interpretable as {\it extended} Weyl algebras. 

We again emphasize that ${\cal H}^{aff}$ is {\it not} simply the tensor product of the CCR with the
rotational (lattice) $C^*$-algebra, but rather the lattice acts on the localization
function via boundary condition.

Within this functorial calculus one
can form diagonal generators in the $k$-fold tensor product:
\begin{equation}\label{22}
W^{(k)} =W^{ext} (\underline f) \otimes ...\otimes W^{ext} (\underline f)
\end{equation}
They generate a subalgebra ${\cal A}^{ext,(k)}_L$ of the $k$-fold tensor product 

It is well-known that the level one $SU(N)$ loop group representation is generated
by the extended Weyl algebra for a particular
lattice $L$ (the root lattice of $SU(N))$. Let us therefore reinterpret the
above tensor-product formula for those special group lattices as:
\begin{equation}\label{23}
\hat W^{(k)} =W^{(1)} (\underline g)\otimes ... W^{(1)} (\underline g)
\quad{\rm with} ~\underline g \in\mbox{ loop-group.}
\end{equation}
with $W^{(1)} (\underline g)$ = loop group representation of level 1.
Then $W^{(k)} \subset \hat W^{(k)}$.

By {\it reduction} we now could construct higher level loop vacuum representation which
are known to lead to plektonic (nonabelian) nets. But such an ad hoc procedure
invoking the loop group and its associated net is not quite in the spirit of
algebraic QFT. I believe that at least for nets with a finite number of
superselection sectors (``rational" chiral theories) there exists a natural plektonic extension of
the tensor product algebra generated by the $W^{(k)}$'s in formula
(\ref{22}). My conjecture is based
on the observation that the loop group nets and the $W$-nets are the only
irreducible plektonic families which implement the classified statistics
families, a remark which generalizes the structural
discussion of plektonic 4-point functions in \cite{37}.
 
We apologize to the reader for these complicated technicalities. They were
only meant to support the hope that chiral conformal QFT can be classified by
purely intrinsic QFT principles (valid in {\it any} dimension) without
reference to particular algebras as Kac-Moody or $W$-algebras. Such messages
are important if one believes that the main physical role of chiral theories
is that of a theoretical laboratory for general QFT.
Algebraic QFT is designed to cope with nets which are far away from the standard
CCR and CAR situations (and their perturbative interactions). The latter can
be considered as function spaces with a nonabelian $C^*$-structure i.e. as
(localizable) functors of Hilbertspaces into von Neumann algebras. Algebraic
nets are supposed to cover the area {\it beyond} such localizable function algebras
i.e. situations in which the locality principle is not implemented through
functions but in a more noncommutative fashion. But in order to explicitly construct examples beyond CCR and CAR
one has to start from the latter because these are the only nets which we
thoroughly know.

\section{Constructive Attempts for $d=2+1$ Anyons}

The most natural idea for the construction of ``free" fields with braid group
statistics is the adaption of Wigner's representation theory of the Poincar\'e
group to this problem. Since the ``little group" is the abelian $U(1)$, the Wigner
wave function are one-component functions on the mass hyperboloid  which transform
under Lorentz-transformations as
\begin{equation}\label{24}
(U(\Lambda)\psi)(p)=e^{is\phi_W(p,\Lambda)} \psi(\Lambda^{-1}p)
\end{equation}
where the Wigner phase $\phi_W(p,\Lambda)$ is a computable real function which, 
as a result of ${\bf R} \subset \widetilde{SO} (2,1)$ represents an element of
the covering of the Lorentz-group. In case of charged particles one has to double
the wave-function space in order to obtain a representation of TCP. 

In the next
step  the Wigner theory has to be extended by a localization concept which was
not  known to Wigner (i.e. different from the Newton-Wigner localization). One
considers real subspaces ${\cal H}_R$ of ${\cal H}$, similar to those of 
chiral current algebra of the previous section. The relevant concept of causally
disjoint localization is then defined  in terms of the symplectic complement using
the symplectic form:
\begin{equation}\label{25}
\sigma (f,g) =Im(f,g)_{\cal H} ,~~~~~ {\cal H} ={\cal H}_W\oplus 
{\cal H}_W^{anti}
\end{equation}
Without local fields it is generally very difficult to explicitly find 
these local subspaces
${\cal H}_R(O)$ which are associated to regions (example: double cones) $\cal O$ in
$d=2+1$ Minkowski space. 

However there is one exception: the (Rindler or
Bisognano-Wichmann) ``wedge"  regions. For wedges the modular Tomita-Takesaki theory 
becomes geometric. On the Wigner space this theory admits a kind of ``pre" 
(or ``first-quantized")-version \cite{38}. Instead of the modular conjugation $J$ and the
modular group $\Delta^{it}$ operators in Fockspace, we have operators $j$
and $\delta^{it}$ acting geometrically on the Wigner wave function space.
The pre-Tomita operator 
\begin{equation}\label{26}
s=j\delta^{1/2}
\end{equation}
is an unbounded closable involutive $(s^2=1$) operator on $\cal H$ \cite{38}
which serves to define
a real subspace ${\cal H}_R$ (wedge) as the $(-1)$ eigenspace of $s$
\begin{equation}\label{27}
{\cal H}_R(wedge)= \mbox{closure~of~real~linear~span}~\{\psi\in{\cal H}~~,~~ 
s\psi = -s \}
\end{equation}
The $s$ is explicitly known, since $\delta$ can be computed from the Lorentz-boost
(which conserves the wedge) and $j$ is obtained from the TCP operator $\vartheta$
on ${\cal H}$ by splitting off a $\pi$-rotation around the $x$-axis (in case of
the standard $t-x$ wedge). 

In the standard $d=3+1$ Wigner theory, this idea 
was recently used \cite{39} to define local nets in a direct fashion, i.e. without local
fields as intermediaries. Since double cones can be obtained as intersection 
of wedges, the double cone algebras can be defined in terms of intersections of
wedge algebras and the isotonic and causal structure of this so defined net 
follows without reference to local fields \cite{39}. 
It is very interesting to extend these ideas to $d=2+1$ anyonic spin representations
\cite{40}.
In this case the statistical phase belonging to the Wigner spin $s \not=$ integer shows
up as a mismatch between the $j$-transformed real subspace (the symplectic
complement) and the geometric opposite wedge (obtained by a spatial rotation):
\begin{equation}\label{28}
j \cdot{\cal H}_R(\mbox{wedge}) \not= {\cal H}_R(\mbox{opp.wedge})
\end{equation}
The operator which removes this mismatch is related to a Klein-transformation. In case of e.g.
${\bf Z}_N$-anyons this Klein-transformation $T$ is again 
 a suitable square root of the $2\pi$-rotational in Theorem 4.

A significant difference between the (semi)integer spin case and genuine
fractional spin appears, if one investigates\cite{40} the possibility of having sharper
localization than just wedge localization. It turns out that for the fractional
case no compact localization is possible, i.e. certain intersection of ${\cal H}_R$
(wedge)-spaces are empty. In those cases one expects the semi-infinite spacelike
string localization to be the best possibility. Therefore one looks for string-like
localized fields. 

The standard way to obtain fields from the Wigner wave
functions is to ``factorize" Wigner's wave functions into covariant wave
functions times a generalization of the $u$ and $v$ spinors. Candidates for
covariant wave functions in $d=2+1$ were first proposed by Mund and Schrader
\cite{41} \cite{23}.
\begin{equation}\label{29}
\psi_{cov}(p.g):=F(L^{-1} (p)g)\psi_w(p)
\end{equation}
with $F$ a (yet unspecified) kinematic function on the covering group $\widetilde{SO}(2,1)$
which has the following equivariance law with respect to the little group:
\begin{equation}\label{30}
F(w\cdot g) =e^{isw} F(g)~~,~~w \in{\bf R} \subset \widetilde{S(2,1)} 
\end{equation}
Indeed this equivariance leads to the covariant transformation
\begin{equation}\label{31}
(U(g)\psi_{cov.}) (p,g') = \psi_{cov} (g^{-1}p,gg').
\end{equation}
Introducing momentum space creation and annihilation operator and covariant operators
in $x$-space according to 
\begin{equation}\label{32}
\psi(x,g)=\int {d^2p\over 2w} (e^{ipx} a(p,g)+e^{-ipx} a^* 
(p,g))
\end{equation}
we have:
\begin{equation}\label{33}
U(g')\psi(x,g) U^+ (g') =\psi (\Lambda(g')x,g'g)
\end{equation}
and
\begin{equation}\label{34}
(\Omega,\psi(x_1,f_1)\psi (x_2,g_2)\Omega)=\int {d^2p\over 2w}
e^{ip(x_1-x_2)} \bar F(L^{-1} (p)g_1)\cdot F(L^{-1}(p)g_2)
\end{equation}
Choosing $x_2$ and $g_2$ ``opposite" to $x_1,g_1$ i.e. 
\begin{eqnarray}\label{35}
x_2 &=& R(\pi)x_1 ,~ R(\pi)~ \mbox{rotation~by}~\pi\nonumber\\
g_2 &=& R(\pi)g_1
\end{eqnarray}
one obtains the statistics factor in agreement with our previous wedge localization:
\begin{equation}\label{36}
(\Omega,\psi(x_1,g_1)\psi(x_2,g_2)\Omega)=e^{2\pi is}
(\Omega,\psi (x_2,g_2)\psi(x_1,g_1)\Omega).
\end{equation}
In order to obtain fields which have a string-like localization, one has to 
specify appropriate functions $F$. As Fredenhagen and Gaberdiel \cite{23} showed, the 
following specification leads to lightlike strings.
\begin{equation}\label{37}
F(g) =e^{ig(0)s}
\end{equation}
with $g(u) \subset{\bf R}$ for $u\in{\bf R} = \tilde S^1 \subset \widetilde{SO(2,1)}
~~$ and $g$ acting in the natural way as a fractional $SO(2,1)$ transformation:
\begin{eqnarray}\label{38}
g&=& (\gamma, w)\nonumber\\
(\gamma,w)(u) &=& w + arg {e^{iu} +\gamma \over 1+e^{-iu}\bar \gamma}
\end{eqnarray}
These lightlike strings should be the singular limit of more natural space-like
strings, but a detailed investigation of the letter is still missing.
Presently it is not clear, if the previous two-point function which was obtained
from the Wigner theory belongs to a spacelike cone localized QFT. But even if
the two point function of ``free" anyons has a support beyond the mass shell,
the modular localization method should reveal why this is so.

A precise comparison with the results obtained on the basis of a Chern-Simons
picture (i.e. a Chern-Simons vector potential interacting with spinor free 
matter)
is difficult because the Chern Simons approach has not been perused far enough 
in order to obtain operators fulfilling braid
group statistics. However on a very formal level it appears that the modified
 spinor matter fields are of the following type \cite{42}
\begin{equation}\label{39}
\psi =: e^{bilin.(\psi_0,\psi_0^+)}  \psi_0 :
\end{equation}
This is similar to the formula of the order parameter in the $d=1+1$ massive Ising
field theory. Such a structure would have a more complicated combinatorics
than the previous ``Wigner-anyons". Whereas the latter are {\it elementary} objects,
the Chern-Simons anyons  are {\it composite} represented in the Fermion Fockspace.
 I presently have no comment on this
discrepancy, except the obvious remark that Chern-Simons anyons are perhaps not
the simplest possible realization of $d=2+1$ braid group statics.

The problem of constructing genuine $d=2+1$ ``free" plektons by algebraic methods
 is at least as
complicated as that ``anyonization" of conformal plektons of the previous
section and there are presently no constructive
ideas.
This area of research promises to be interesting, lively and controversial for
some time to come.

\section{Concluding Remarks}

In this work we argued that algebraic QFT presently plays its most constructive
role in $d\le 2+1$ QFT's with plektonic statistics were Lagrangian method fail.

For $d=3+1$ QFT's the algebraic method suggests two areas of potential progress:
a reinvestigation of perturbative problems in the sense of section 4,
including an intrinsic understanding of ``interaction" independent of the use
of special ``field coordinates", and a better
understanding of the physical principles behind gauge theories.

Algebraic QFT always maintained a critical distance to an alleged ``local gauge
principle". Therefore there is no particular reason for an algebraic field
theorist to react towards some new paradoxical looking situations (e.g. the
alleged clash between the Seiberg-Witten duality and the ``local gauge 
principle"\cite{15}) with surprise.
 To the contrary, a critical evaluation from an algebraic
(rather than a differential geometric) standpoint would lead one to believe
that among all attempts to press a new physical reality into the quasiclassical
straight-jacket of Lagrangian quantization, local gauge theories is  the
most useful one. The formal gauge idea suggests a new
potentially fruitful concepts of algebraic QFT: the enhancement of 
degrees of freedom  in asymptotic limit theories (``asymptotic 
freedom'') \cite{16}.

Earlier attempts to ``problematize" the notion of magnetic field 
in terms of operator 2-cohomology on the same
level of depth as the problematization of the notion of charge in the theory
of superselection sectors (which can be reinterpreted as local
 operator 1-cohomology\cite{39}) have not met with
much success \cite{39}. In the light of recent observations on the 
e.m. duality in supersymmetric QFT's, the main problem from the algebraic point of view
is the understanding of a local physical property of the net generated by the
field strength $F_{\mu\nu}$ which excludes the ``free phase", i.e. 
 excludes the simultaneous absence of
electric as well as magnetic charges. The free electromagnetic field has,
as we have seen, a
peculiar obstruction, not shared by other massive free fields. It violates
Haag duality for topologically nontrivial regions. It is believed
that $F_{\mu v}$'s with nontrivial charges may not show this obstruction.
 For dual order-disorder situations in
$d\le 2+1$ theories such a criterion would not be necessary, since the case
that both dual variables leave the vacuum sector invariant (i.e. do not
create charges) is automatically excluded. But a classification of superselection
rules coming from Maxwellian-like e. and m. charges remains difficult,
  since the bad infrared behaviour prevents a localization of charges in
arbitrarily thin  semi-infinite spacelike cones.

The lesser popularity of algebraic QFT as compared to the geometric quantization
approaches seems to be a result of its conservative attitude with respect to
inventions.

However underneath the conservative surface of the algebraic method hides a revolutionary new
way of looking at QFT, a change of paradigm with respect to Lagrangian 
quantization. In more philosophical terms, the latter complies with the
Newtonian way of understanding physical reality: a space-time manifold with
a material content. The net point of view of algebraic QFT with its relations and inclusions
between isomorphic local algebras (hyperfinite $III_1$ von Neumann factors)
on the other hand
harmonizes much more with Leibniz view of reality as coming about through
the relations of monades ($\simeq$ the hyperfinite $III_1$ von Neumann 
factors representing the local algebras).
This is of course also the underlying philosophy about mathematics in  the Jones subfactor theory.

Looking at present QFT-related theoretical issues, it is not difficult to find
a rough division into three topics: Renormalization group QFT, String theory
and Algebraic QFT. 

The first one has led to a philosophy \cite{44} that one should
be happy with cutoff Lagrangians.  The new aspect of the philosophy underlying
the renormalization group in QFT is not that it states that reality
is like an ``infinite onion", each layer has its principles only to be
 superseded by the next one. 
Physics was always like this, and old principles were always eventually rendered
limiting cases of new ones.  Rather it resides in the message that to strife
for conceptual and mathematical consistency in each layer is 
not required any more, since our lack of knowledge about the next layer can be
summarily taken care of by cutoffs (even though this violates our {\it present}
principles). With other words the principles with which this century
had such a glorious start are an illusion, ``effective" theories with cutoffs
are the new
 goal. This ideology looks modest, apart from its tendency to label ``reductionist" 
viewpoints (as the one in this article)
as scientific arrogance [44, page 157]. 
To me it appears as the physicists analogue of
Fujikawas socio-economical ideology.

The second topic, namely string theory will have to make a big effort in order
not to end as the ``mathematically most useful physical phlogiston theory of the
$20^{th}$ century" \cite{55}. 
\footnote{
With the following qualifications added, I partially share this 
pessimistic outlook.
Although the e.m.-duality concepts are touching upon very deep 
physical issues, I think that the global quasiclassical methods as 
well as the more recent differential- and algebro-geometric 
consistency approach are limited to some very special
global features of supersymmetric duality without any presently visible
relation to the problem of e.m.-duality in $d > 2+1$ local quantum
physics. In this article I alluded to the latter problem in terms of 
the more noncommittal terminology as the "problematization of the
magnetic field" on the same level of conceptual profoundness as the 
successfull problematization of "charge" in the theory of 
superselection sectors.

In a very optimistic point of view, the aforementioned global attempts
parallel the global Kramers Wannier low-dimensional order-disorder
duality and the local understanding can be hoped for as arising e.g. 
from progress in zero mass localization concepts in QFT (telling us 
why we should go beyond the semiinfinite space like string localization
of the first section.)

But looking at the present sociology in physics it is very difficult to entertain 
such optimism. The fact that string theory has created a new theoretical realm which 
can be understood without profound knowledge of the development of quantum 
field theory may be the reason why it attracts so many young physicists 
and has led to a globalization of the market, with its cleansing effects on long 
term intellectual investments. 
One may be afraid that there is little room left for thinkers in the "Bohr 
Heisenbergian" spirit who are prepared to make that big investment 
which could reconquer the lost conceptual terrain.} 
Rampant mathematics unbridled by physical principles and concepts does  not seem
to be the way out of the most profound crisis. 

Areas close to ones own research one usually criticizes more
mildly. However it is not a secret, that algebraic field theorist are a bit like
 religious preachers. Their everyday life (teaching courses on QFT by using
a 40 year old formalism) is often quite apart from those high
conceptual values they preach.  My hope is that this will change in the near future.
Algebraic QFT is attractive to me, because it remained a ``family" activity and
one does not feel compelled to make
unfulfillable claims.

There is an interesting historical precedent that a conservative approach, i.e.
one which is faithful to physical principles but revolutionary in their
implementation may at the end be the more successful
 one. I am again referring to the discovery of renormalization theory in 
1948/49 which did not use any of the many inventions of those days but 
just consisted in an elaboration of a conservative ideas and the pursuit of its
intrinsic logic. 
For a very informative account I refer to  Weinberg's new textbook \cite {22}.\\
\\
{\bf\large Acknowledgements}

Most of this work was carried out at the Universidade Federal do 
Espirito Santo, Brazil. I am indebted to Prof. J.C. Fabris for the 
invitation and to my colleagues at the UFES for their helpful 
attitude. I also feel obliged to thank for the hospitality at the ESI in Vienna 
where I enjoyed many stimulating discussions about the the content of 
this work in particular with Prof. R. Stora.


\begin{thebibliography}{99}
\bibitem{55} The author of this aphorism did not want to be credited.
\end{thebibliography}
\end{document}